\documentclass[10pt]{article}

\usepackage{graphicx}
\usepackage{subfigure}
\usepackage{color}
\usepackage{times,amsfonts,enumerate,amssymb,amsmath,epsfig,subfigure,bm,cite}
\usepackage{amssymb}

\begin{document}

\date{}

\title{FAST, PARALLEL AND SECURE CRYPTOGRAPHY ALGORITHM USING LORENZ'S ATTRACTOR}

\author{
ANDERSON GON\c{C}ALVES MARCO\\
Universidade de S\~{a}o Paulo\\  
Instituto de Ci\^{e}ncias Matem\'{a}ticas e de  Computa\c{c}\~{a}o \\ 
S\~{a}o Carlos, S\~{a}o Paulo, Brazil   \\
lock@grad.icmc.usp.br \\
\\
ALEXANDRE SOUTO MARTINEZ \\
Universidade de S\~ao Paulo\\ 
Faculdade de Filosofia Ci\^encias e Letras de Ribeir\~{a}o Preto \\
Ribeir\~{a}o Preto, SP, Brazil and \\
National Institute of Science and Technology in Complex Systems \\
asmartinez@usp.br \\
\\
ODEMIR MARTINEZ BRUNO \\
Universidade de S\~{a}o Paulo\\
Instituto de F\'{i}sica de S\~{a}o Carlos  \\ 
S\~{a}o Carlos, S\~{a}o Paulo, Brazil \\
bruno@ifsc.usp.br
}

\maketitle

\begin{abstract}
A novel cryptography method based on the Lorenz's attractor chaotic system is presented. 
The proposed algorithm is secure and fast, making it practical for general use. 
We introduce the chaotic operation mode, which provides an interaction among the password, message and a chaotic system. 
It ensures that the algorithm yields a secure codification, even if the nature of the chaotic system is known. 
The algorithm has been implemented in two versions: one sequential and slow and the other, parallel and fast. 
Our algorithm assures the integrity of the ciphertext (we know if it has been altered, which is not assured by traditional algorithms) and consequently its authenticity. 
Numerical experiments are presented, discussed and  show the behavior of the method in terms of security and performance. 
The fast version of the algorithm has a performance comparable to AES, a popular cryptography program used commercially nowadays, but it is more secure, which makes it immediately suitable for general purpose cryptography applications. 
An internet page has been set up, which enables the readers to test the algorithm and also to try to break into the cipher in.
 
\end{abstract}

\section{Introduction}

Since ancient times, the art of encrypting messages using ciphers or breaking into them, has decided the destiny of the civilization. 
Throughout the history, in many situations, cryptography was directly responsible for the fortune or doom of kings and queens. 
The success of exchanging secret messages decided the future of many battles and even wars. 
Some of them were important and very famous, such as the Thermopiles battle (Greece - 490 BC), where according to the legend, the Spartans were notified of the Persian invasion by a crypto message hidden in earthenware, which enabled the Spartans to plan a defense strategy before the Persian attack. 
Although the secret of the information plays an essential role in military strategy, it was only during the XX century wars that its real importance become obvious and popular. 
With the advent of radio, it was possible to exchange vital strategy information by electromagnetic waves. 
Since there are no boundaries for radio waves, the cryptography power appeared as a fundamental strategy element.  
It is responsible for ensuring that the enemy, although being able to receive the message, is not capable of understanding it. 
In fact, cryptography was decisive in the destiny of the first and second world war. 
Although cryptography is an important military and political information tool, these uses overcame the war times. In the digital ages, cryptography ensures security of business and bank transactions over the internet, and it can also ensure privacy of private messages, making the modern computer, the internet and wireless networks secure and usable technologies. 
 
The objective of cryptography consists of encrypting a message, file, document or image.
To understand this procedure, consider the following definitions. 
One thinks of the object to be encrypted as a string (vector) of size $n_p$. 
The components of this vector is transformed to decimal ASCII representation (integers ranging from $0$ to $2^8-1$) forming the \emph{plaintext}, which is represented by $\vec{P}$, with size $n_p$. 
Also, one needs a \emph{password}  string, which is transformed to decimal $ASCII$ representation forming $\vec{\pi}$, with size $n_{\pi}$. 
The plaintext $\vec{P}$ is transformed into the ciphertext $\vec{C}$ via  an \emph{encode function} $\vec{C}= E_{\vec{\pi}}(\vec{P})$, which is parametrized by the password $\vec{\pi}$. 
To retrieve the plaintext from the cipher, one uses the \emph{decode function} $\vec{P} = D_{\vec{\pi}}(\vec{C})$, where $D_{\vec{\pi}} = E^{-1}_{\vec{\pi}}$ is the inverse of the encode function and parametrized by $\vec{\pi}$. 
Perhaps the simplest encode function is the monoalphabetic (Cesar) cipher $E_{\pi}(P_i) = (P_i + \pi) \mod 2^8$, where $2^8$ is the length of our alphabet, $P_i$ is a component of $\vec{P}$ and $\pi$ is the integer associated to the password. 
The associated decode function is: $D_{\pi}(C_i) = (C_i - \pi) \mod 2^8$. 
%[r1
Here, we call attention to  the reversibility property of the $\mod$ operation. 
Care must be paid because the signal in one of the arguments of the $\mod$ operation is reversed, for instance,  $C =  A+B \mod D$  and $A = C- B \mod D$.

The current cryptography methods are based on discrete mathematics (symmetrical key)~\cite{serpent,towfish,idea,aes} and number theory (asymmetrical key)~\cite{logaritimo,RSA,curvas}. 
These algorithms use elementary mathematical as ciphers and have to increase the computing complexity to ensure a secure crypto. 
Taking this into account, they carry out many bit-to-bit operations and permutation between neighboring elements, or they have to use very long keys (asymmetrical algorithms - nowadays RSA algorithms, for instance, use keys varying between $2^{10}$ and $2^{11}$ bits) to hide information. 
However, for both strategies, the increase in the computing complexity and the size of the keys only ensure that the cipher is secure for a certain time.  
The constant increase in computer power and the evolution of the cryptoanalyses algorithm have made it easy to break into. 
Nowadays, although there are algorithms which are capable to breaking into this kind of ciphers, they take too long, making them impractical ~\cite{crip_linear,crip_diferenca,criptoanalise1}.  
The increase in computational power makes algorithms based on elementary mathematics weak. 
One example is the DES method~\cite{crip_diferenca}, which was popular and largely used in the past. 
However, the algorithm became too weak and is no longer in use nowadays. 

To make the ciphers more secure, some new mathematical methodologies have to be considered and consequently new cryptography approaches arise. 
In this context, the chaos theory and the complex dynamical systems appear as very attractive alternatives to develop cryptography methods~\cite{wang_2008}. 
Chaotic systems have some characteristics that make them valuable for cryptography, such as: 
(i) complex numerical patterns, 
(ii) unpredictably for unknown initial conditions, 
(iii) strong dependence on the initial conditions, 
(iv) based on relatively simple equations and 
(iv) determinism. 
The cryptography algorithms based on chaos generally explore the symmetrical approach. 

Cryptography algorithms using chaos appeared in the 90's. 
The first ones shuffle continue time signals using chaos~\cite{Pecora:1990,Chua,Parlitz:1996p969,Bu:2004p1023} and also digital signals~\cite{Murali}.  
At the end of the decade, chaos cryptography applied to digital data was introduced. 
In 1998, Baptista~\cite{Baptista} proposed the use of the logistic map as a pseudo random generator to make a cipher capable of dealing with bit sequences. The Baptista's algorithm was improved by Wong and collaborators~\cite{Wong:2001} making it faster and more secure. 
The block cryptograph was incorporated in the logistic map in 2005 by Xiang {\em et. al.}~\cite{blocos} as an enhancement of the previous methods. 
As far as we are aware, there are two proposals in the literature using the Lorenz's attractor as basis of cryptography methods. 
The first one~\cite{PRE} was proposed in 1998, but it is very simple and weak.
In the second one~\cite{AliPacha:2007}, the Lorenz's attractor is used to generate pseudo random numbers in the same sense as the logistic map based ciphers. 

In this work a novel cryptography method based on the Lorenz's attractor is introduced.  
The proposals of the algorithm are to be secure and fast, making it practical for general use. 
In cryptography theory, the strongest cipher (impossible to break) is reached when the pad has the same length as the message (OTP - One-time pad cipher)~\cite{Bell_Labs}. 
We use the Lorenz's attractor to compute a virtually infinite pad exploring the chaos. 
The complexity of the attractor and its determinism, makes it possible to compute chaotic and long number sequences. 
What makes the OTP impossible to be broken in is the fact that one does not know the pad. 
For instance, if one uses a book to write a OTP cipher, the cipher is decoded when the book is discovered.
Since we use the Lorenz's attractor to compute a pad, although it is difficult, one can compute the Lorenz's trajectory and discover the cipher pad.  
To make this task harder to perform, we have added an operation mode in chaos, which we call the chaotic operation mode. 
The chaotic operation mode is a technique, inspired by traditional cryptography (operation mode~\cite{modo_operecao}), which provides an iteration between the password, already coded message and chaotic system to make a complicated walk over the attractor. 

The presentation of this paper starts with a discussion and a description of the chaotic operation mode (Section~\ref{operation}). 
The cryptographic algorithm is proposed in details in Section~\ref{algorithm} and the results are showed the Section~\ref{results}, illustrating the behavior of the proposed method. 
In Section~\ref{conclusion}, the paper ends with a discussion about the method.     

\section{Chaotic Operation Mode}
\label{operation}

Cipher-block algorithms are able to carry out bit-to-bit operations between the neighbors of the plaintext and the password. 
This approach has a disadvantage of making similar ciphertext sequences for similar blocks of the plaintext, consequently making a weak cipher. 
The operation mode was created to correct this fault, which main idea is to make the blocks independent. 
Consequently, the ciphertext is encrypted combining the information of the password plus the previous plaintext sequence. 
It can ensure that similar blocks of the plaintext are codified as different ciphertext, making the cipher stronger.

One of the most popular operation mode algorithms is Cipher-Block Chaining (CBC)~\cite{modo_operecao}. 
Each component of the plaintext $\vec{P}$, of size $n_p$ is \emph{xored} with the previous ciphered element: $C_{i}=E_{\vec{\pi}}(P_{i}\oplus C_{i-1})$, where $C_{0}$ is an arbitrary value used to encode the first component of $\vec{C}$ that can also be used as an extra password. 
This value has an important role in the crypt process, since without it one cannot decode the message. 
The value of $C_0$ is generated from the password and it is desirable to include an additional parameter which may be different for each user. 
The symbol $\oplus$ represents the bitwise XOR logical operation. % [r1] we added the word "bitwise" to be more precise.    
We notice that the size of $\vec{C}$ is $n_c = n_p + 1$ and that the elements of $\vec{P}$ and $\vec{C}$ are integers ranging from $0$ to $2^8 -1$ representing the decimal ASCII code. 
Here, the encode function $E_{\vec{\pi}}$  represents any symmetrical cryptography algorithm~\cite{serpent,towfish,idea,aes} using password $\vec{\pi}$. 
The operation mode uses information from the previous coded sequence ($ C_{i-1}$) to code $C_{i}$ to ensure that similar sequences in $\vec{P}$ do not have the same sequences in $\vec{C}$. 
The inverse process also combines previous coded sequence to decode: $P_{i}=D_{\vec{\pi}}(C_{i} \oplus C_{i-1})$, where $D_{\vec{\pi}}$ is the decode function.

%[r1] added paragraph
We observe that encryption and decryption operation can carried out by the same procedure due to the reversible property of the bitwise XOR logical operator. 
To illustrate the reversibility property, consider two bit sequences $\vec{P}$ and $\vec{\pi}$ and the resulting bit sequence $\vec{C} = \vec{P} \oplus \vec{\pi} = \vec{\pi} \oplus \vec{P}$. 
One then has $\vec{P} = \vec{C} \oplus \vec{\pi} = \vec{\pi} \oplus \vec{C}$ and $\vec{\pi} = \vec{P} \oplus \vec{C} = \vec{C} \oplus \vec{P}$.
% We would like to comment that the bitwise XOR logical operator is also used in other crytographic schemes. 
There are other operators that allow the reverse operation, for instance, as mentioned in the introduction, the $\bmod$ combined with a sign inversion in the argument. 
One interesting cryptographic scheme uses cellular automata~\cite{oliveira_2004}. 
In this case, the cellular automata must be reversible to preserve the information of the initial states~\cite{mora_2002,mora_2003,hernandez_2003}. 
For a cellular automaton to be reversible, the global function must be bijective to have an inverse.  
Local functions that involve only bitwise XOR logic operation lead to a linear global function, therefore, a  bijective function (with non-singular transition matrix)~\cite{delrey_2006,delrey_2009}.

Since the preserved patterns of the plaintext in the ciphertext may give clues for cipher analysis and the cipherblock chain is not strong enough, here, we propose to use dynamical systems in the chaotic regime to accomplish this task. 
We call this procedure of chaotic operation mode.   
				
The chaotic operation mode has the same basis of the traditional operation mode. 
The message, password and previous code are combined to achieve a strong cryptography not preserving repetitive patterns of the original message. 
While the operation mode combines the message and the previous coded message using elementary operations, the chaotic operation mode uses a chaotic system to compute this combination. 
This approach improves the cryptography, making the analysis of it more difficult. 

In the following, we show how to use a dynamical system in cryptography. 
Firstly consider $n_l = n_c$ vectors $\vec{r}$, with initial condition $\vec{r}_0$. 
%[r1] text added 
Notice that $\vec{r}$ is a vector, where each component represents a variable of the Lorenz attractor so that it is a three dimensional vector. 
Each  vector is obtained by the iteration: $\vec{r}_{i} = \vec{F}[\vec{r}_{i-1} + \vec{M}(P_{i})]$, where $\vec{F}(\vec{x})$ is a given vectorial function of a vector $\vec{x}$ which  represents the dynamical system. 
The components of the vectorial function $\vec{M} = (m_1,m_2,m_3)$, with $0 \le m_i \le 0.255$, with $i=1,2,3$, transforms the argument $P_i$ (an integer between 0 and 255) depending on the iteration step.
Notice that the plaintext characters alter step by step the iteration process, which in a parameter region of chaos gives rise to the chaotic operation mode. 
The elements of the ciphertext are obtained from: $C_{i}=E_{\vec{\pi}}[ P_{i} + f(\vec{r}_{i-1} )]$, which encrypts the scalar $P_i$, using bit-to-bit operations with bits extracted from the vectors $\vec{r}$. 
The function $f(\vec{r})$ extracts a value from the vector $\vec{r}$ of the dynamic system. 
The decode process is simply achieved with the inverse operation: $P_{i}=D_{\vec{\pi}}[C_{i} + f(\vec{r}_{i-1})]$.  

Recently a security fault in the logistic map based algorithm that could be extended to other chaos system was demonstrated~\cite{Arroyo}. 
It is based on the return of the map that can be explored to break into the cipher. 
The chaotic operation mode solves this security fault of the chaos based cryptography algorithms, since it combines the password and plaintext with the chaos system, making it more secure.     

When using the chaotic operation mode we call attention to the fact that if a message is altered we are able to detect it. 
In this way we guarantee its authentication.  
Besides, we can also guarantee the integrity of cipher. 
The chaotic operation mode can guarantee the authentication because a  minimal change in $C_{i}$ leads to a huge modification in the Lorenz's path, that can be easy identified, assuring integrity and consequently its authenticity.  

\section{Lorenz-based Cryptography Algorithm}
\label{algorithm}

We propose a full method that combines the dynamic system $\vec{F}(\vec{x})$ with the encode function $E_{\vec{\pi}}$. 
In our case, the dynamical system is the Lorenz's attractor that generates a trajectory which is combined with the message, password with the chaotic operation mode. 
We start defining the quantities used in the algorithm and then we carefully describe it.

\subsection{Definitions}
\label{definitions}

The proposed algorithm uses the Euler differential discretization method to numerically compute the Lorenz's equation system (other numerical methods could be considered) $\vec{r}_{i+1}= \vec{F}(\vec{r}_i)$, with $\vec{r} = (x,y,z)$ and:
	\begin{eqnarray}
\nonumber	
	x_{i+1} & = & [(1 - \sigma) x_{i} + \sigma y_{i} ] \delta t \\
\nonumber
	y_{i+1} & = & ( \rho x_i - 2 y_{i}  - x_i z_i  ) \delta t \\
	z_{i+1} & = & [x_i y_i + (1 - \beta) z_{i} ] \delta t \; ,	
      \label{eq:lorenz_attractor}
	\end{eqnarray}	
The parameters $0< \delta t \leq 0.027$, $\sigma =10.0$ $ \rho =28.0$ and $\beta= 8/3$ are values where the system gives rise to chaotic sequences. 
The number of iterations $n_{it}$, along the Lorenz's attractor, determines the speed of the processing of the algorithm. 
Here we have used $n_{it} = 3000$.
		
The password $\vec{\pi}$ is converted to decimal ASCII values, with size $n_{\pi}$. 
Notice that $n_{\pi}$ has a limit,  this is because of the representation of a real number in double precision in a given computer. 
The quantity  $\xi = 52$, according to the IEEE754 convention, is the number of bits of the mantissa to represent a double precision number. 

The first stage of the algorithm is devoted to the conversion of the password into a coordinate of the Lorenz's attractor space. 
The vector $\vec{\pi}$, of size  $n_{\pi}$, is converted into a vector $\vec{a} = (a_1,a_2,a_3)$, with the following components:
\begin{eqnarray}
				a_1 & = & \left\{ \begin{array}{ll} 
					\sum_{i=1}^{L} \pi_{i}2^{8(i-1)}                &\textrm{if} \; n_{\pi} \bmod 3 = 0\\  
					\sum_{i=1}^{L} \pi_{i}2^{8i}+\pi_{3L+1}  &\textrm{if} \; n_{\pi} \bmod 3 \neq 0\\
				\end{array} \right.\\
				a_2 & =& \left\{ \begin{array}{ll} 
					\sum_{i=L+1}^{2L} \pi_{i} 2^{8[-L+(i-1)]}                 & \textrm{if} \; n_{\pi} \bmod 3 \neq 2\\  \\
					\sum_{i=L+1}^{2L} \pi_{i}2^{8(-L+i)} + \pi_{3L+2} & \textrm{if} \; n_{\pi} \bmod 3 = 2\\  
				\end{array} \right. \\
				a_3 & = & \sum_{i=2L+1}^{3L} \pi_{i}*2^{8[-2L+(i-1)]} \; ,
\end{eqnarray}
where $L = \mbox{floor}(n_{\pi}/3)$.
The vector $\vec{a}$ is then converted to $\vec{a}'= (g(a_1), g(a_2),g(a_3))$, where $g(x)=x/10^{\mbox{ceil}(\log[2^{8(L+1)}]) }$ so that $0 \le a'_i \le 1$, with $i=1,2,3$. 

The starting point to move in the Lorenz's attractor is written as:  
\begin{equation}
\vec{r}_0= \vec{a}' + \vec{\lambda} \; , 
\end{equation}
with $\vec{\lambda} = (\lambda_1,\lambda_2,\lambda_3)$. 
The components of $\vec{\lambda}$ are chosen to have values in the following range: $ -15.67 < \lambda_1 <16.01$, $-11.28 < \lambda_2 <16.01$ and $0.090 < \lambda_3 <62.000$. 
These ranges guarantee stable chaotic phases.  

\begin{figure}[!htb]	 
	\centering
	\includegraphics[width=0.4\columnwidth]{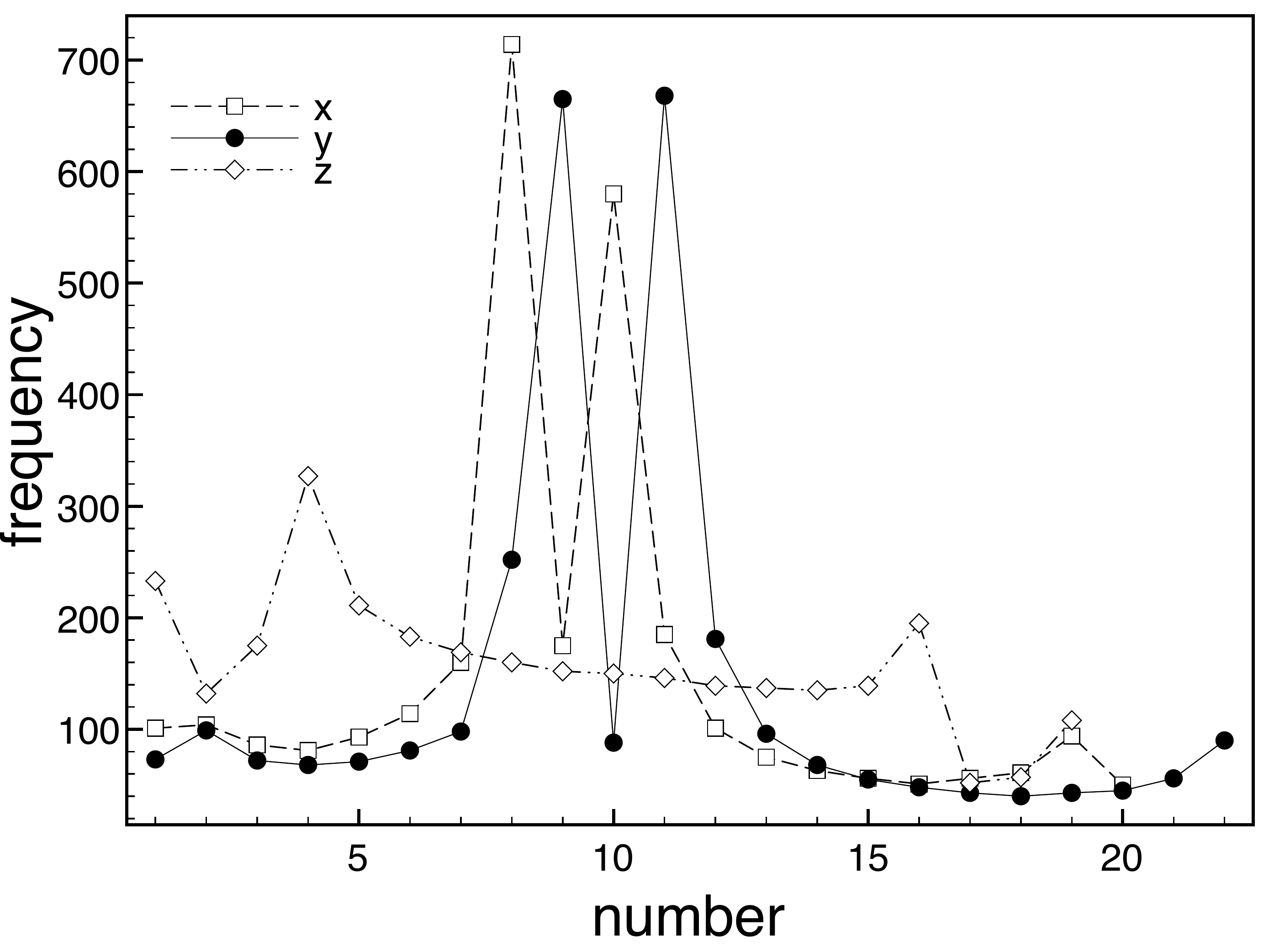}(a)\\
	\includegraphics[width=0.4\columnwidth]{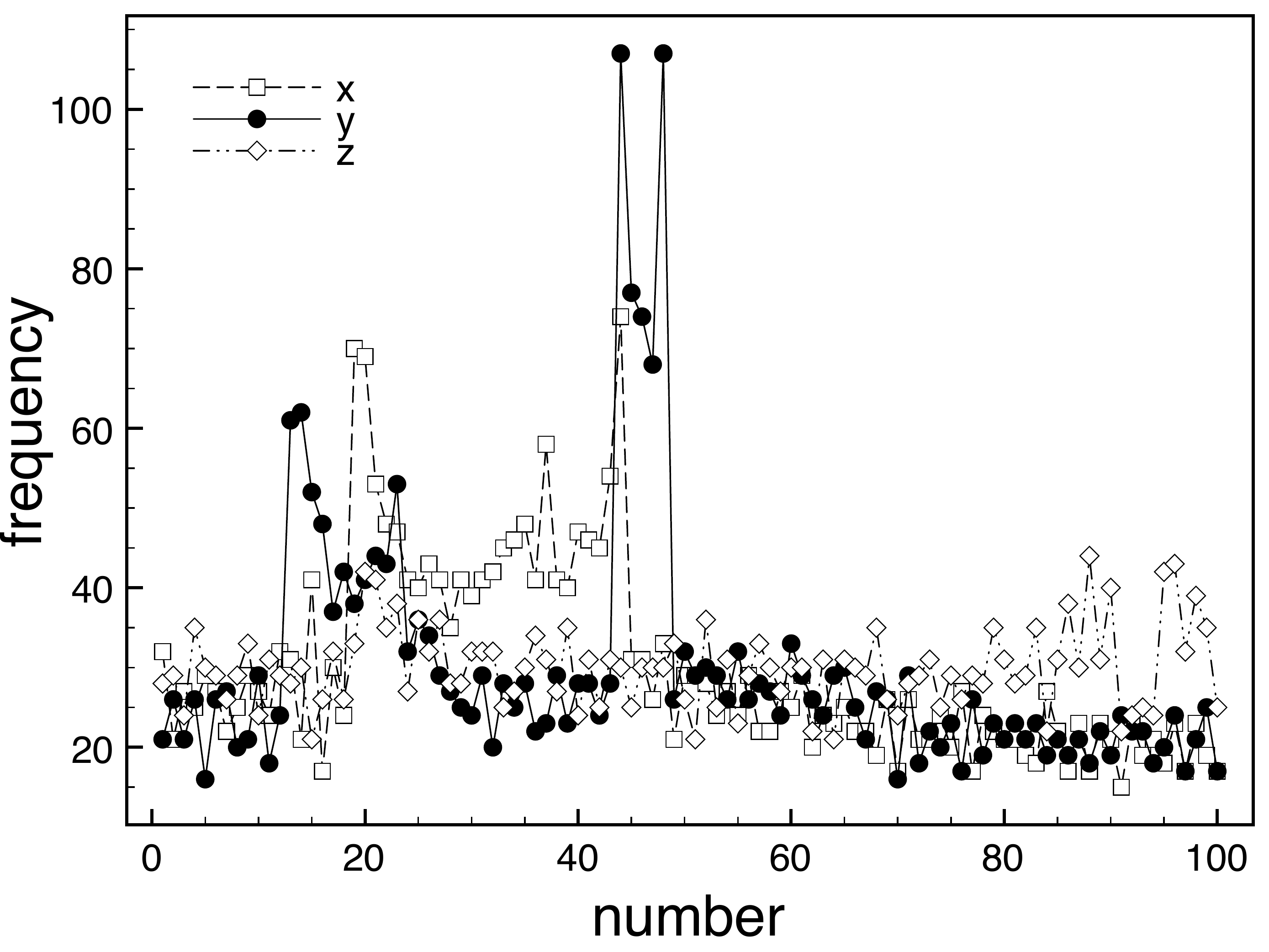}(b)\\
	\includegraphics[width=0.4\columnwidth]{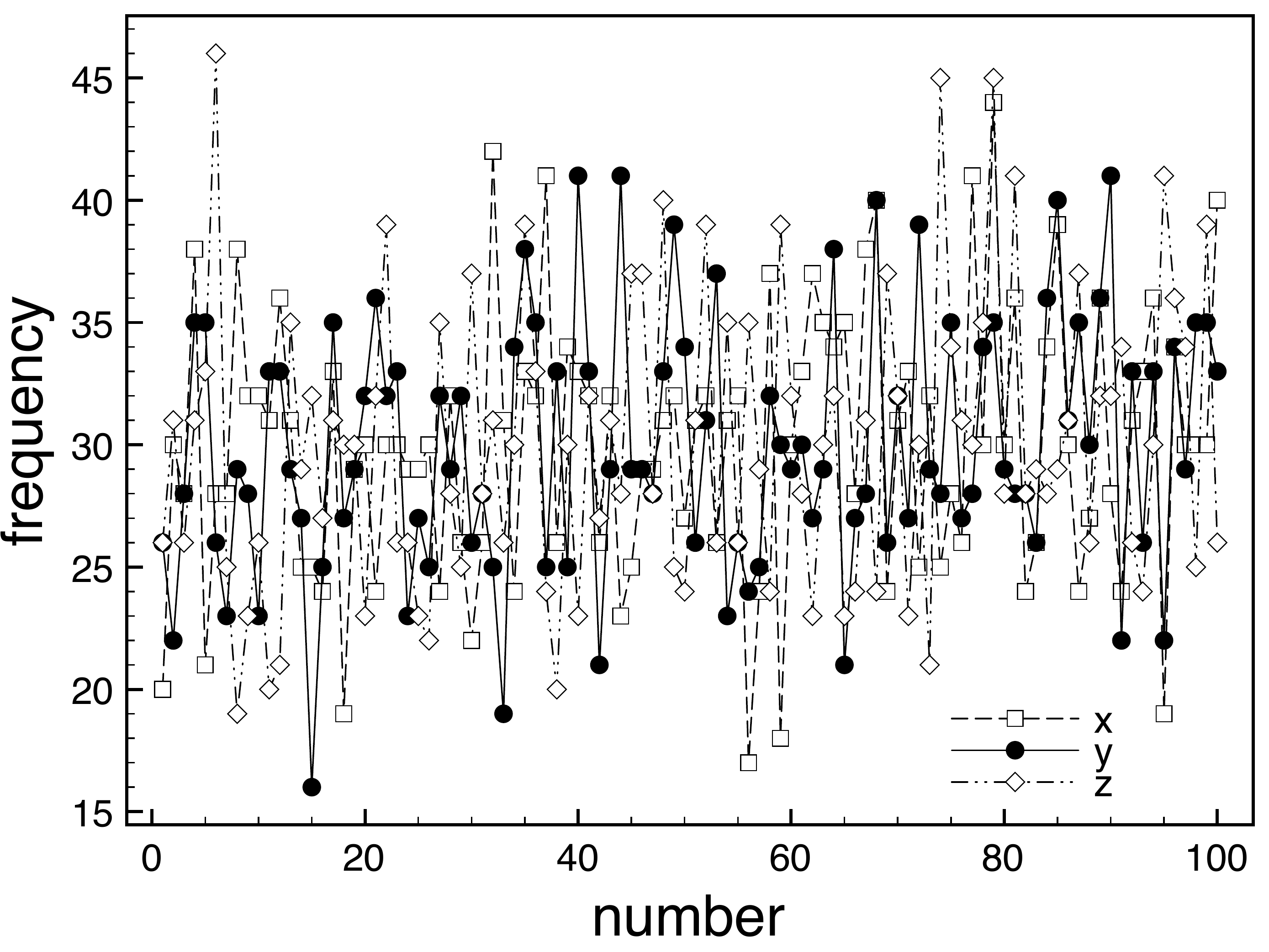}(c)\\
	\includegraphics[width=0.4\columnwidth]{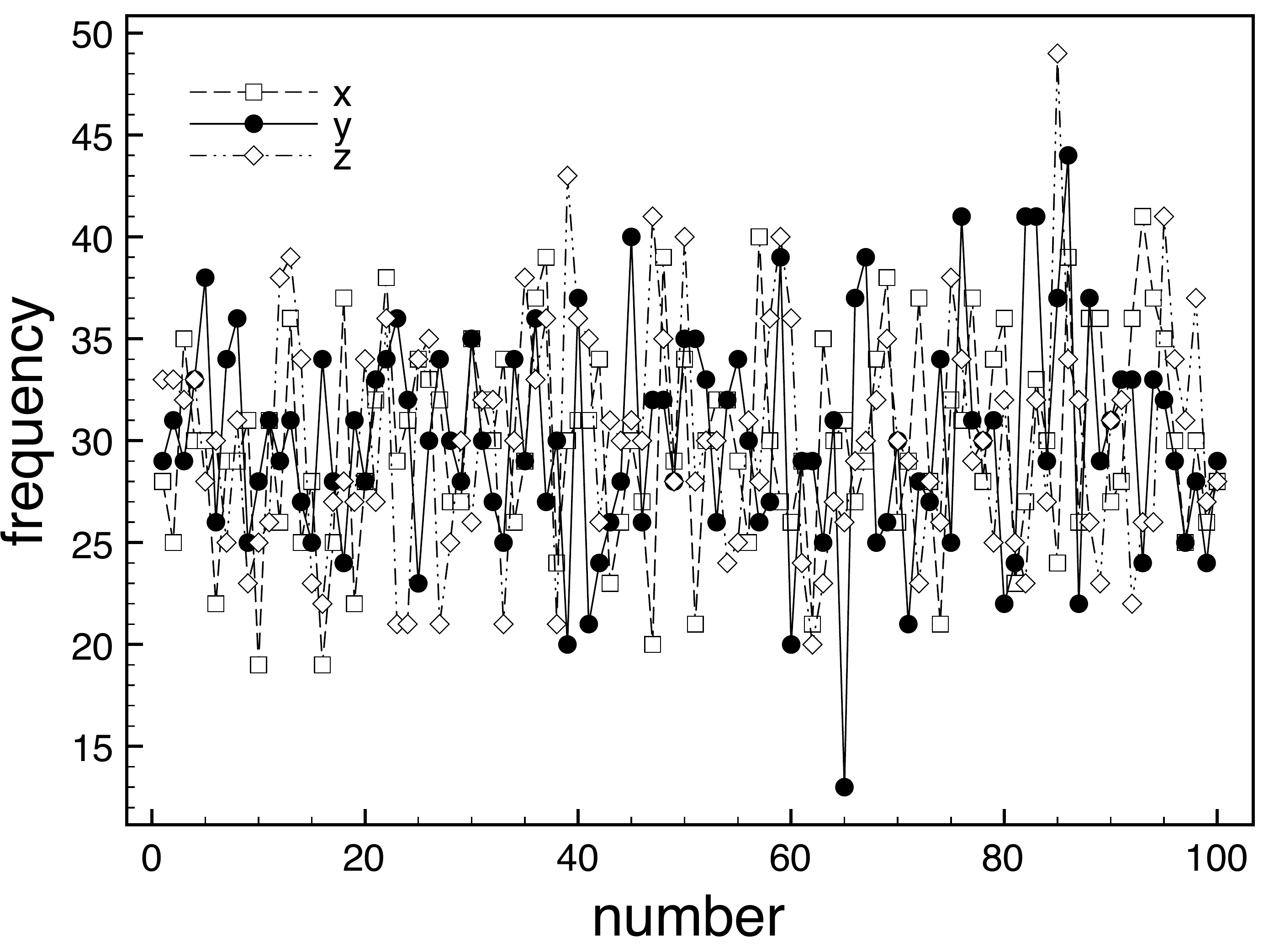}(d)
	\caption{Frequency distribution of the integer part as well as the decimal part of the Lorenz's  attractor. 
	(a) integer part, 
	(b) first two digits of the decimal part,  
	(c) third and fourth digits of the decimal part and 
	(d) fifty and sith digits of the decimal part.
	 }
	\label{fig:number_frequency}
\end{figure}

Firstly, compute $\vec{\mu} =(\mu_{1},\mu_{2},\mu_{3})$, so that: $\mu_{1} = (a_1 + a_2 + a_3) \bmod 3$, $\mu_{2} = (a_1 * a_2 + a_3) \bmod 3$ and $\mu_{3} = (a_1 + a_2 * a_3) \bmod 3$.  
Compute now, $\vec{\alpha} = (\alpha_1,\alpha_2,\alpha_3)$, so that:  
			\begin{equation}       
				\alpha_{i}=\left\{ \begin{array}{ll} 
					x_{n} & \textrm{, } \mu_{i}= 0 \\  
					y_{n} & \textrm{, } \mu_{i}= 1 \\  
					z_{n} & \textrm{, } \mu_{i}= 2  
				\end{array} \right.\; ,
				\label{eq:alpha1}
		       \end{equation}	
with $n=0$.
Now, choose values for $\vec{k} = (k_1,k_2, k_3)$, with the components of this vector in the range:  $2 < k_i \leq \mbox{floor}[(\xi-14)/8]$, so that $k_i$ being an integer. 
Finally, compute $\vec{\Omega} = (\Omega_1,\Omega_2,\Omega_3)$ so that: 
\begin{eqnarray}
\Omega_{i} & = & \text{hash}(i \vec{a}) \bmod k_{i}  \; , 
\end{eqnarray} 
where $\text{hash}(\vec{a})$ is a hash function such as MD5 or SHA. 
These values have been chosen to obtain the less significant bytes in the coordinates of the Lorenz's attractor.
This contrasts to the procedure presented in Ref.~\cite{PRE} and Ref.~\cite{AliPacha:2007} where the integer and first digits from the decimal part are considered. 
The procedure we propose makes the cipher frequency distribution uniform not giving clues for undesirable decodification as show in Sec.~\ref{results}. 
In Figure~\ref{fig:number_frequency} we depict the frequency distribution of the integer part as well as the decimal part of the Lorenz's attractor. 
In Fig.~\ref{fig:number_frequency}(a), we show that the integer part of coordinates of the Lorenz's attractor are concentrated in the middle of the graph varing in a short range. 
In Fig.~\ref{fig:number_frequency}(b), we show that the first digits of the decimal part of the coordinates of the Lorenz's attractor are concentrated. 
In Fig.~\ref{fig:number_frequency}(c) and (d), we show that the frequency distribution is practially uniform for the 3rd and 4th and 5th and 6th pairs of digits. 
We call attention to the fact that the low significant digits are the heard of our algorithm. 

To encrypt a plaintext, we propose a new encode function.
This function combines an integer (a decimal ASCII of the plaintext) and a float point (obtained from the trajectory in the Lorenz's attractor) in the argument to produce an integer, between 0 and 255, to represent the coded character. 
 For this reason one has a structure as: $E_{\vec{\alpha},\vec{\Omega}}(P_i) = [ P_i + f(\vec{\alpha},\vec{\Omega})] \bmod 2^8$. 
  
Let us now concentrate in $f(\vec{\alpha},\vec{\Omega})$, with $\vec{\alpha}$ and $\vec{\Omega} $ are quantities related to the password $\vec{\pi}$ and depend on the Lorenz's attractor.  
Here we only use the components $\alpha_1$ and $\alpha_2$, which are float point numbers and $\Omega_1$ and $\Omega_2$, which are integer numbers. 
The algorithm to obtain these quantities in the successive iterations is presented in Section~\ref{encdec}, where we show that $\alpha_3$ and $\Omega_3$ are auxiliary quantities.  
 
 Our objective is the creation of a ternary operation, which operates on a character code and on two float point numbers so that an inverse operation exists, combined with the password $\vec{\pi}$, and one can easily perform the decode operation. 
 To accomplish this task, consider a float point number $\alpha$, so that it mantissa has $\xi$ bits and we convert it to an integer multiplying it to $10^{\nu}$, where $\nu= \text{floor}(\log{ 2^{\xi - 6}})$ with the function $\text{floor}(x)$ taking the integer part of $x$. 
The product $\text{floor}(\alpha 10^{\nu})$ is an integer with 16 bytes. 
Recall that the integer 255, is a byte of $1$'s. 
To move this eight $1$'s to the left, one multiplies it by $2$, so that, multiplying $255$ to $2^8$, one displaces 255 of one byte. 
To displace $255$ of $\Omega$ bytes, one must multiply it to $2^{8 \Omega}$.
The number $255 \times 2^{8 \Omega}$ is a mask.  
To obtain the $\Omega$-th byte of the quantity $\text{floor}(\alpha10^{\nu})$, apply the logical AND operation, represented by $\wedge$, with the mask  $255 \times 2^{8 \Omega}$. 
A resulting large float number is obtained dividing the above result by $2^{8 \Omega}$. 
Consider the function: 
\begin{equation}
R_{\nu}(\alpha,\Omega) = \frac{\text{floor}(\alpha10^{\nu}) \wedge (255\times2^{8 \Omega})}{2^{8 \Omega}} \; .
\label{eq_principal_anderson}
\end{equation}

Now we are able to write the encode function as:  
		\begin{equation}
		y= E_{\vec{\alpha},\vec{\Omega}}(x)  =   \left[x+ \sum_{i=1}^2 R_{\nu}(\alpha_i,\Omega_i) \right] \bmod 2^8 \; , 
		\label{eq:encode}
		 \end{equation}
and the function used to decode the information as:
			\begin{equation}
			x = D_{\vec{\alpha},\vec{\Omega}}(y)  =  \left[ y - \sum_{i=1}^2 R_{\nu}(\alpha_i,\Omega_i) \right] \bmod 2^8 \; , 
			\label{eq:decode}
			 \end{equation}

To prepare data for the following steps set: $\vec{\alpha'} =\vec{\alpha}$,  $\vec{\mu'}=\vec{\mu}$ and $\vec{\Omega'}=\vec{\Omega}$.
%[r1
We call attention to the reversibility of the $\mod$ operation of Eqs.~\ref{eq:encode} and~\ref{eq:decode}. 
We have chosen to use the $\bmod$ operator instead of the bitwise $XOR$ operator because its full reversibility asks for an additional operation (inversion of the argument). 
Another important aspect is that the reversibility of $\bmod$ is not so obvious as simple change of the state of bits. 
These two points combined make the codification more difficult to be broken in with $\bmod$.

We have chosen to use the $\bmod$ operator, which gives the rest of a division, instead of the bitwise $XOR$ operator because to have a full reversibility, an additional operation (inversion of the argument) is necessary in $\bmod$ operation. This makes the codification more difficult to be broken in.

\subsection{Successive Iterations}
\label{encdec}

The process of encrypting and decrypting a message can be divided into 3 steps, which are shown below. 
The dynamic system  iterates the components of $\vec{P}$ or $\vec{C}$. 
Some parts of the algorithm are different for the first iteration. 
This is necessary because the method uses variables calculated in the previous iteration. 

\begin{description}
			
\item[Step 1:] The function $E$ (encryption) or $D$ (decryption) is used to compute $P_i$ or $C_i$: $C_i = E_{\vec{\alpha},\vec{\Omega}}(P_i)$, for encrypting and $P_i = D_{\vec{\alpha},\vec{\Omega}}(C_i)$, for decrypting, where $E_{\vec{\alpha},\vec{\Omega}}$ and $D_{\vec{\alpha},\vec{\Omega}}$ are given by Equations~\ref{eq:encode} and~\ref{eq:decode}.
	
\item[Step 2:] The chaotic operation mode is carried out. 
Consider the quantity $\Omega_3$, the password $\vec{\pi}$, and  $\Theta = P_{i}/10^{3 + \Omega_{3}}$. 
The Lorenz's attractor is calculated to shuffle the elements and ensure that there are no possibility of similarity of cipher blocks when there is a similarity in the blocks of the plaintext. 
The chaotic operation mode changes one of the Lorenz's parameters, adding $\Theta$ to the $x$ component if $\mu_3 =0$, to $y$ if $\mu_3 = 1$ or to $z$, if $\mu_3 = 2$.
	
\item[Step 3:] The quantities $\mu_{i}$  are calculated as: $ \mu_{i}' =  [\mu_{i} + R_{\nu}(\alpha_i,\Omega_i)] \bmod 3$  where $R_{\nu}$ is given by Eq.~\ref{eq_principal_anderson}. 
The components of $\vec{\alpha}$ are given by Eq.\ref{eq:alpha1}. 
Consider now $k_3$ an integer in the range $0 < k_3 < \nu - 2$ and calculate $\Omega_{i}  =  [\Omega'_{i}+ R_{\nu}(\alpha_{[(i+2) \bmod 3]+1}, \Omega_i') ] \bmod k_{i}$, where $R_{\nu}$ is given by Eq.~\ref{eq_principal_anderson}. 
Finally set: $\vec{\alpha'} =\vec{\alpha}$,  $\vec{\mu'}=\vec{\mu}$ and $\vec{\Omega'}=\vec{\Omega}$, $\vec{r}_{n} = \vec{r}_{n} + \vec{a}'$.

\end{description}

We stress that this algorithm may be adapted to use other dynamical system that presents chaos as the maps obtained from discrete population models~\cite{martinez_2009}. 
	
\section{Experiments and Results} 
\label{results}

Perhaps, there is no cipher that cannot be broken into, and never will be, at least, using conventional computers and numerical mathematics. 
History demonstrates that, many ciphers, which were considered invincible were in fact broken into. 
Indeed, the cryptography and cryptoanalysis are in constant conflict, in which, there are no permanent winners. 
New ciphers are always appearing and new cryptoanalysis methods emerging to break into them (the red queen effect). 
As a result, it is hard to predict, how difficult it is to break into a cipher. 
Nevertheless, there are some metrics that can help the analysis and can estimate how much the cipher makes the ciphertext shuffled and unintelligible, allowing for a supposition about how strong the cipher is. 

In the following, two experiments with the proposed algorithm are presented. 
Firstly, the algorithm encrypts a text message and an image. 
The analysis of the shuffled capability of the method is presented. 
Besides the security, the computer performance is another important point for a cipher to be able to be used in real world applications. 
Secondly, the results of the performance for the two versions of the sequential and parallel algorithm are presented. 
The results are compared with a popular AES method, which is used commercially nowadays.  

When a ciphertext is yielded, several analysis estimating the strength of the cryptography are made. 
In the text experiment, the book ``The Return of Sherlock Holmes'' was encrypted. 
Long texts, such as books, are easier to be broken into when compared with short messages since statistical patterns may be explored. 

\begin{figure}[!htbp]	 
	\centering
	\includegraphics[width=0.5\columnwidth]{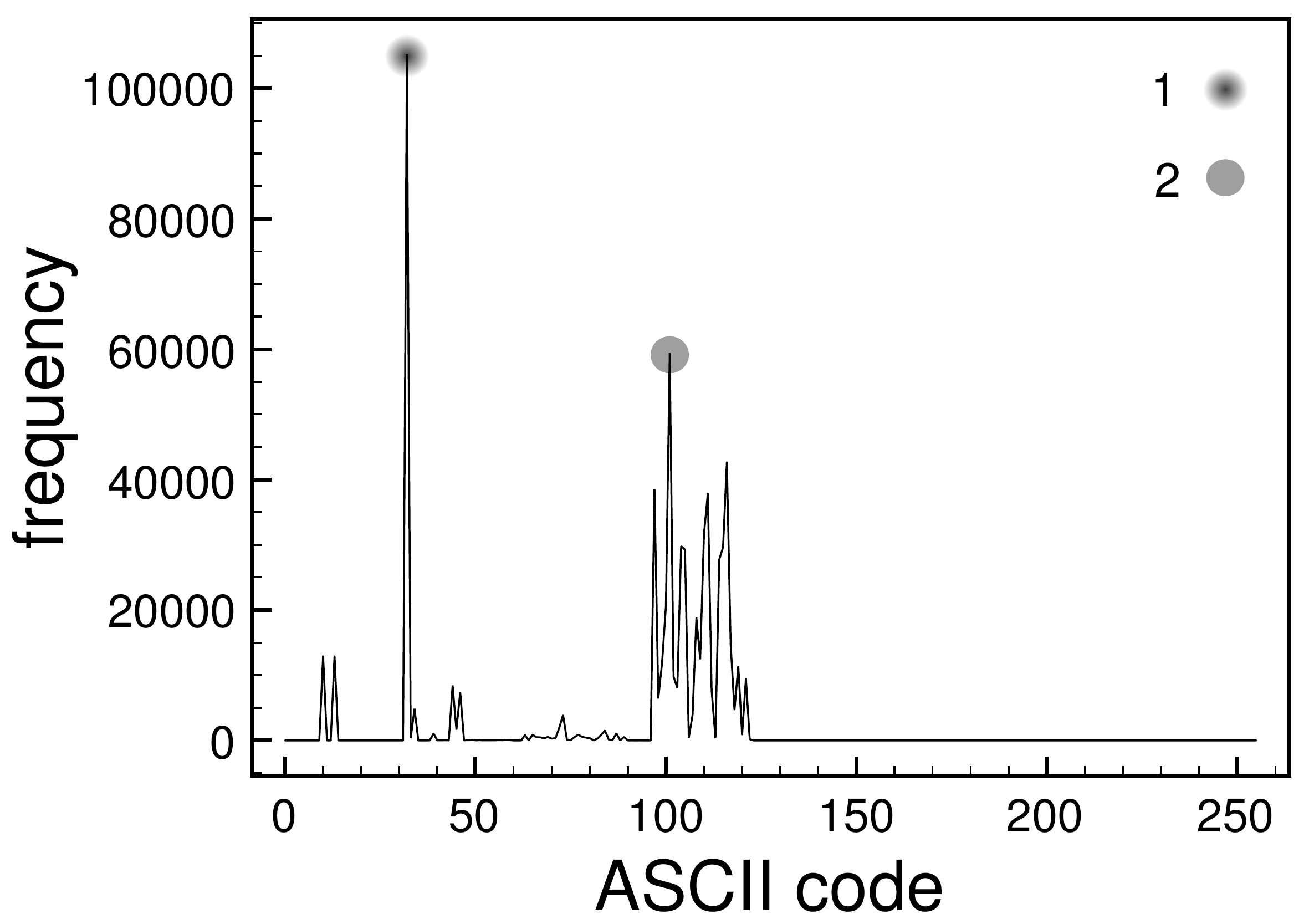} \\
	(a)
\\
	\vspace{0.5 cm}
	\includegraphics[width=0.5\columnwidth]{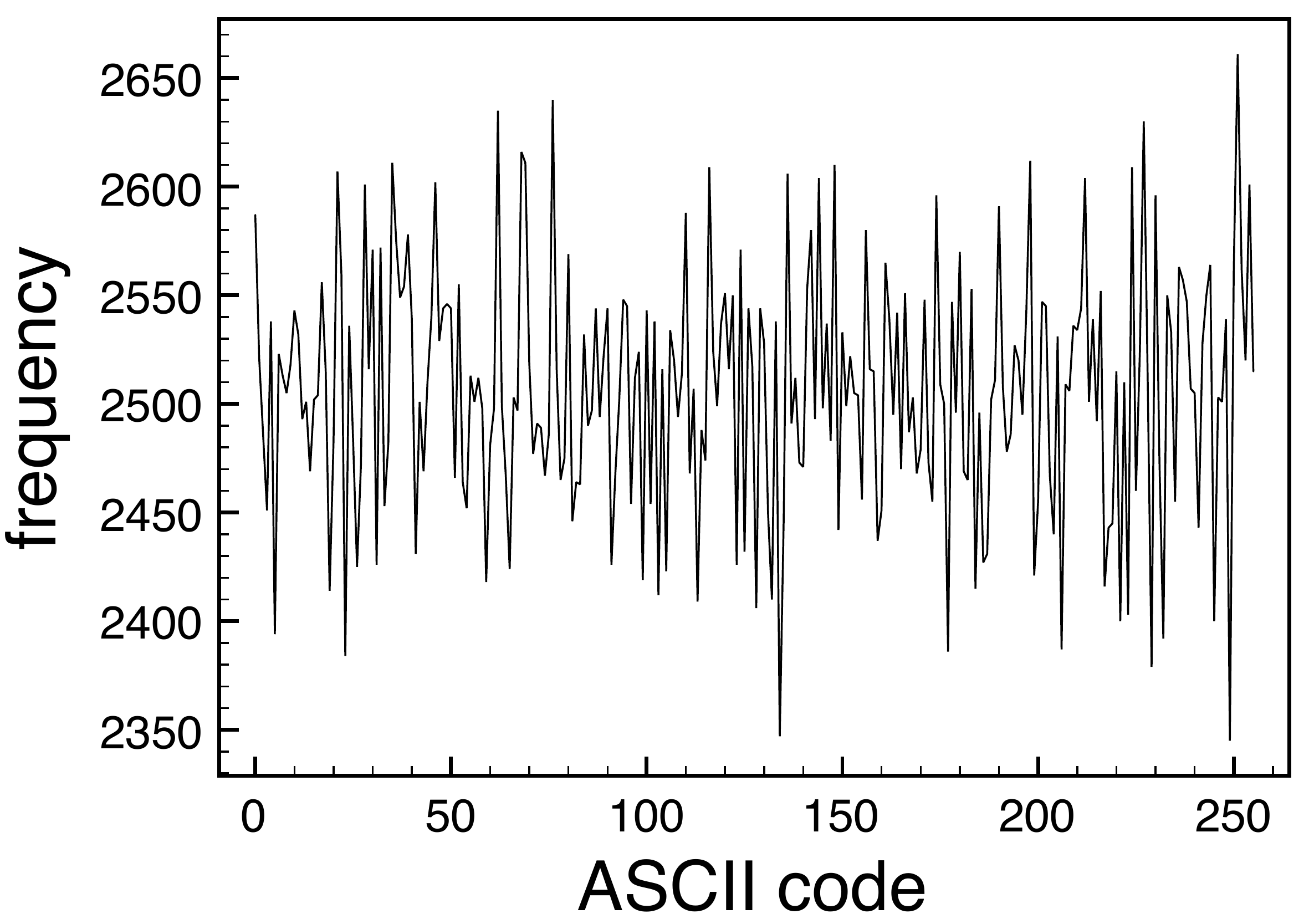}\\
	(b) 
\\
	\vspace{0.5 cm}
	\includegraphics[width=0.5\columnwidth]{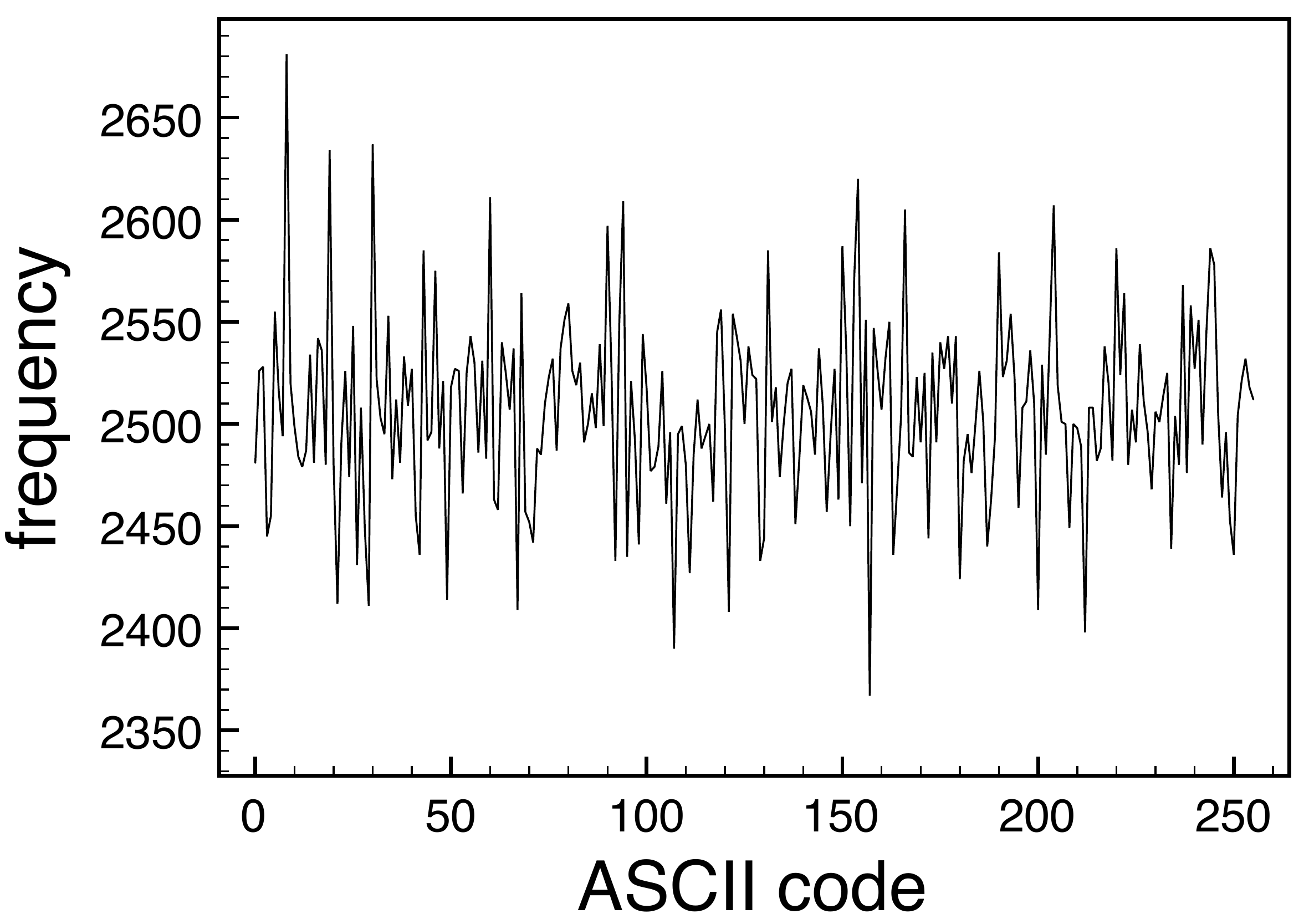}\\
	(c)

	\caption{The frequency analysis of the book ``The Return of Sherlock Holmes'' using slightly different passwords. (a) Histogram of the plaintext the circle 1 represents the frequency of character 'space' and  the circle 2 represents the frequency of character 'e'  , (b) histogram of the ciphertext using ``123456'' as a password and (c) ciphertext histogram using ``123457' as a password.}
	\label{book_histogram}
\end{figure}

One of the most well known methods of cryptoanalysis is the frequency analysis. 
It is quite simple and consists of carrying out a histogram, which presents the ASCII characters frequency, giving tips about their frequency pattern for the cryptoanalysis.    
The frequency analysis is carried out and displayed in Figure~\ref{book_histogram}. 
It illustrates the histogram of the considered book, comparing the plaintext (Figure~\ref{book_histogram} (a)) with the ciphertext (Figure~\ref{book_histogram} (b) and (c)). 
Notice that, the distribution is almost constant  in the ciphertext, like a white noise, which illustrates a good performance for the crypto. 
Figures~\ref{book_histogram} (b) and (c) present the book's histogram encrypted with two slightly different  passwords:``123456'' and ``123457'. 
The two histograms have different plots, illustrating the way the cryptography result depends on the password.

\begin{figure}[!htb]	 
	\centering
	\includegraphics[width=0.7\columnwidth]{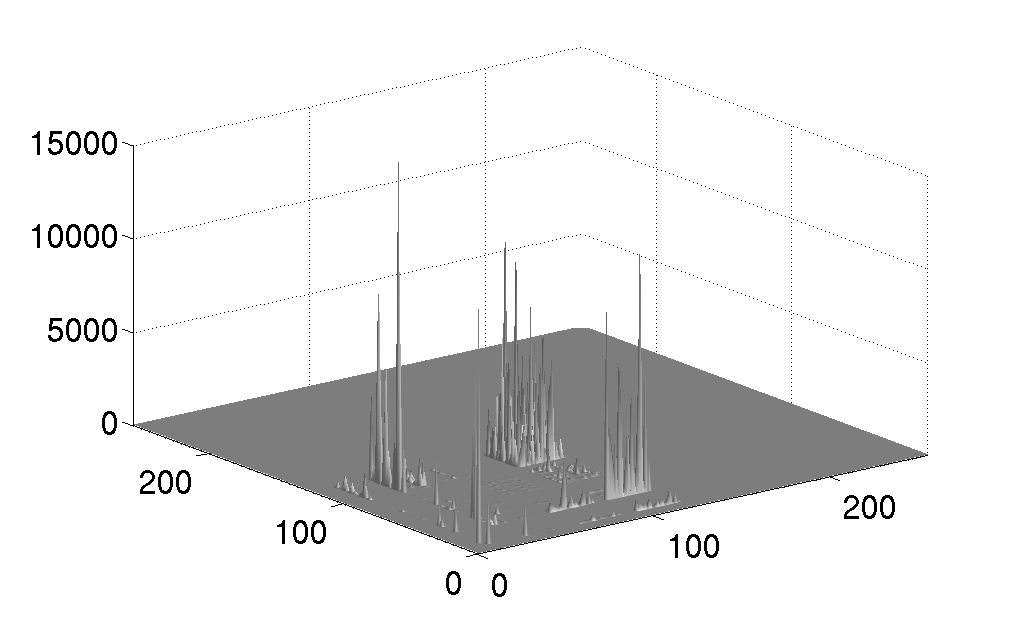} (a)
\\
	\vspace{0.5 cm}
	\includegraphics[width=0.7\columnwidth]{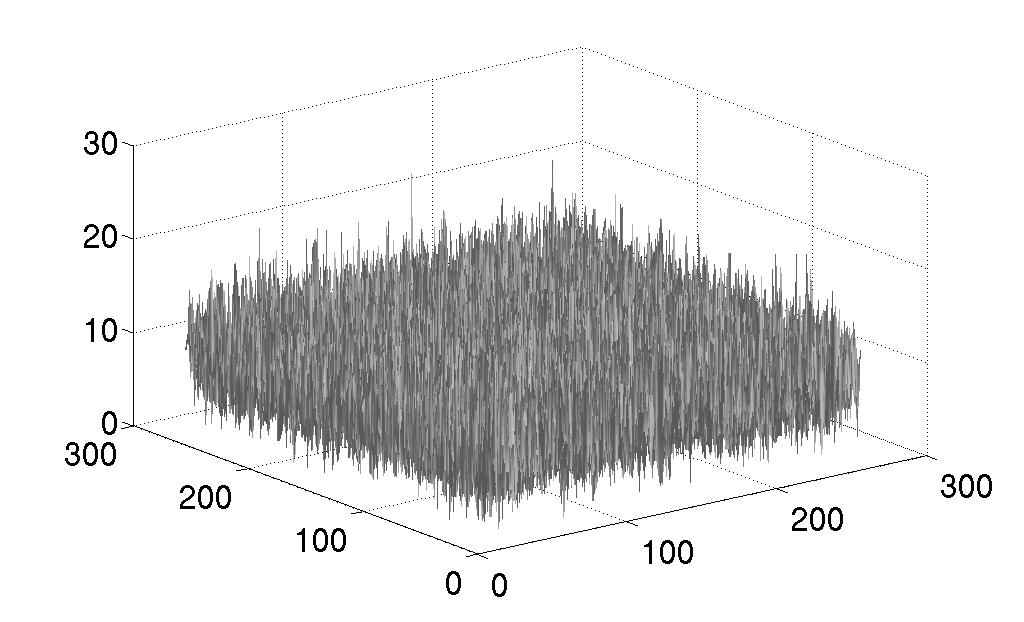}(b) 
	\includegraphics[width=0.7\columnwidth]{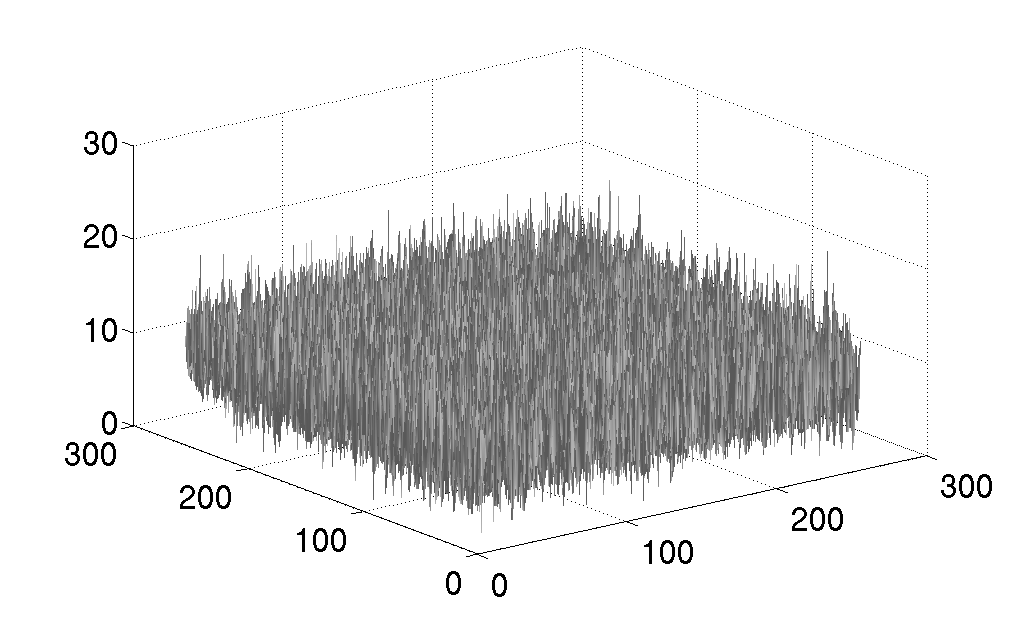}(c)

	\caption{The auto-correlation matrices of the book ``The return of Sherlock Holmes'. (a) Auto-correlation matrix of the plaintext, (b) auto-correlation matrix of the ciphertext and (c) auto-correlation matrix of a white noise signal.}
	\label{book_correlation}
\end{figure}

The Shannon's entropy is another approach to estimates how unintelligible the ciphertext is. 
For a eight bit codification the text entropy is $S = - \sum_{i=0}^{2^8-1} p_{i} \log_{2}(p_{i})$, where $p_i$ is the relative frequency of the $i$-th ASCII character. 
The maximum entropy value is obtained when all the $p_i$ are the same ($p_i = 1;2$) leading to $S_{max} = 8$. 
The entropy of the book plaintext is 4.5 while, the entropy of the ciphertext attains its maximum value $S=S_{max} = 8$. 
The entropy achieved by the cryptography is the same as the theoretical prediction for the system. 
This result shows that the ciphertext has a very low level of redundancy or predictability. 

In Figure~\ref{book_correlation} auto-correlation matrices are shown, comparing the plaintext (Figure~\ref{book_histogram}) (a) with the ciphertext (Figure~\ref{book_histogram} (b) and a white noise signal with the same length of the book (c)).  

\begin{figure}[!htb]		
	\centering
	\includegraphics[width=0.32\columnwidth]{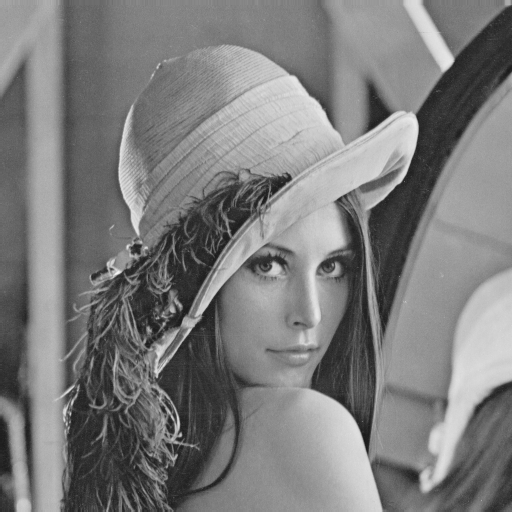}
	 \includegraphics[width=0.32\columnwidth]{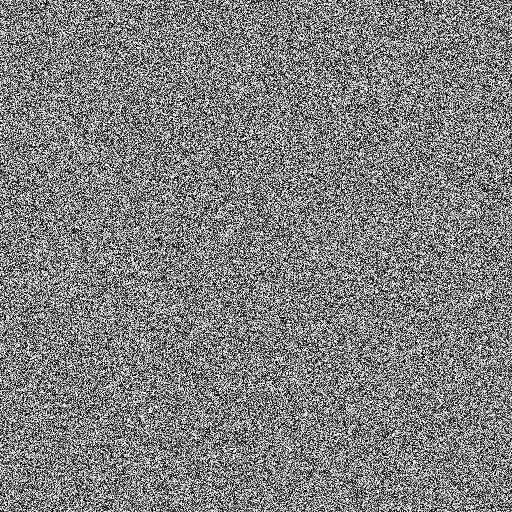} 
	 \includegraphics[width=0.32\columnwidth]{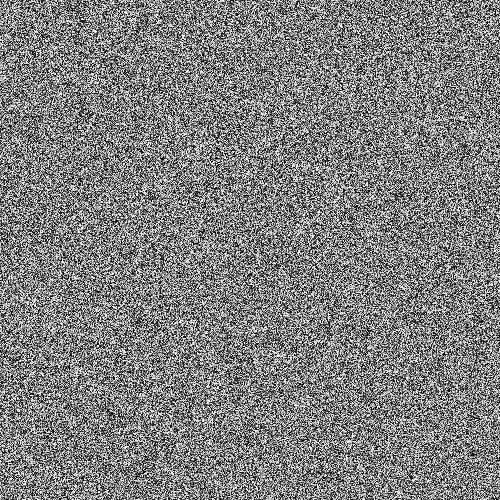}
	 \\
	 (a)\hspace{0.28\columnwidth}(b)\hspace{0.28\columnwidth}(c)
	 \\
	 \vspace{0.3cm}
 \includegraphics[width=0.32\columnwidth]{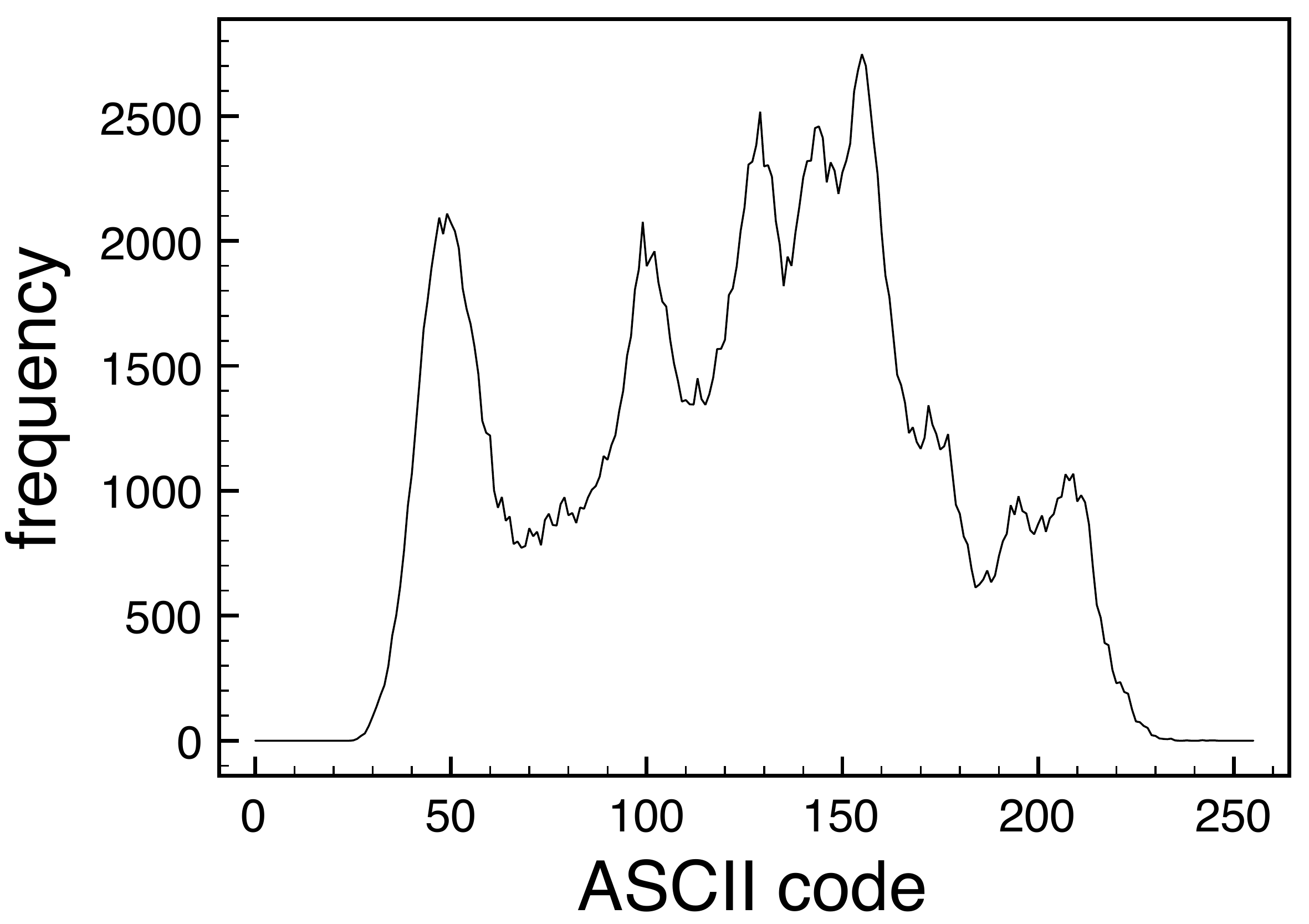}
\includegraphics[width=0.32\columnwidth]{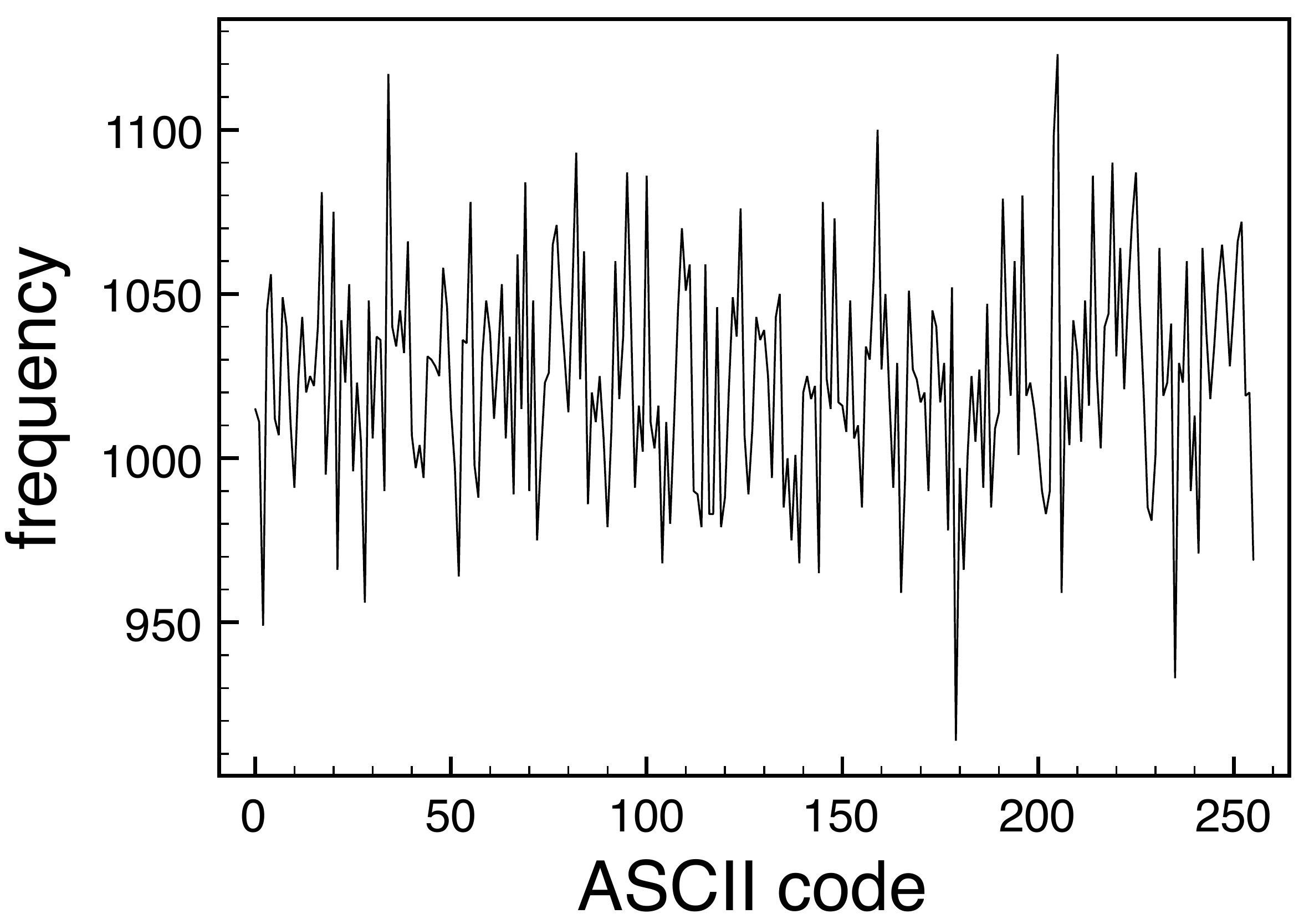} 
\includegraphics[width=0.32\columnwidth]{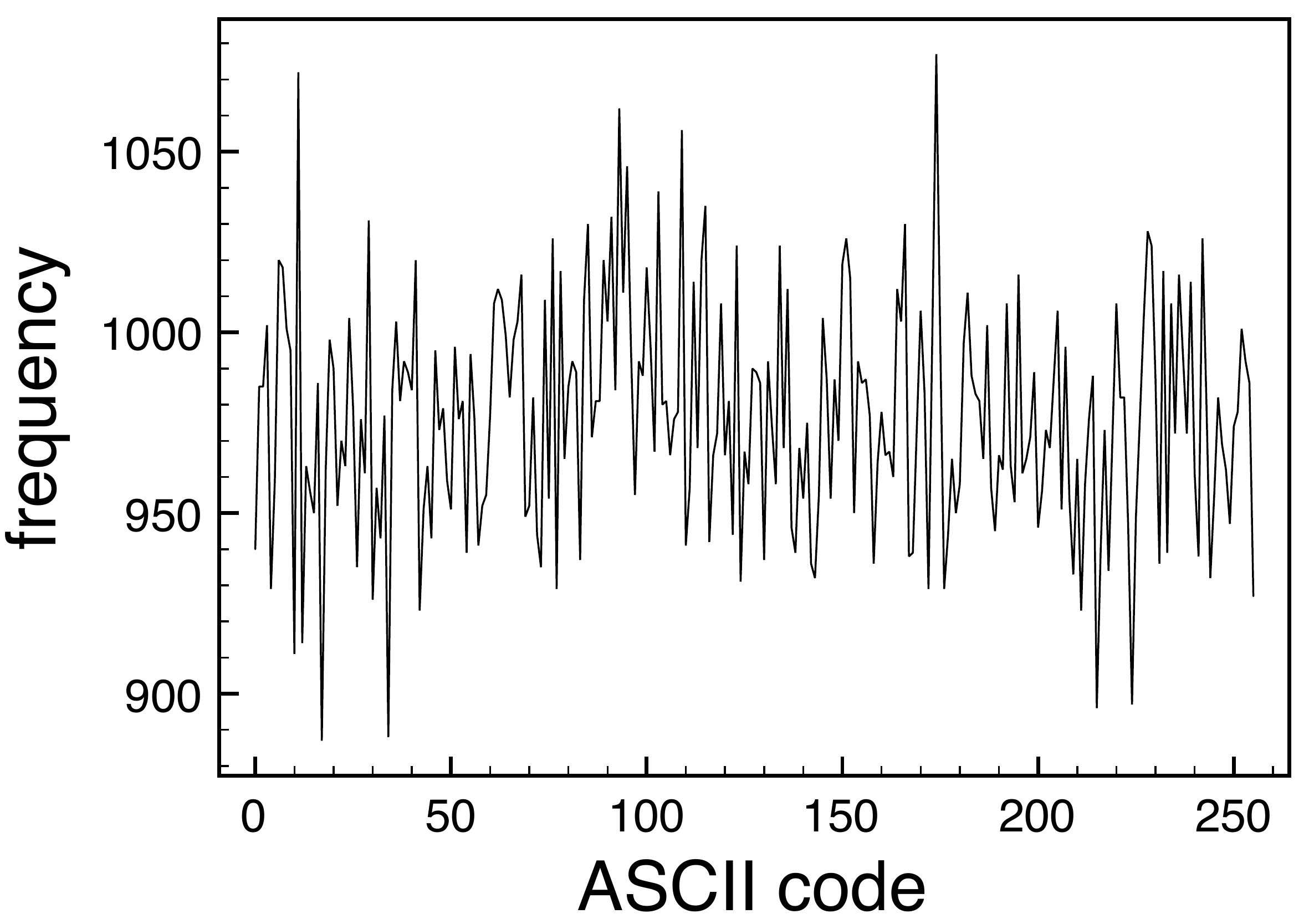}
	 \\
	  (d)\hspace{0.28\columnwidth}(e)\hspace{0.28\columnwidth}(f)
	 \\
	 \vspace{0.3cm}
	 \includegraphics[width=0.32\columnwidth]{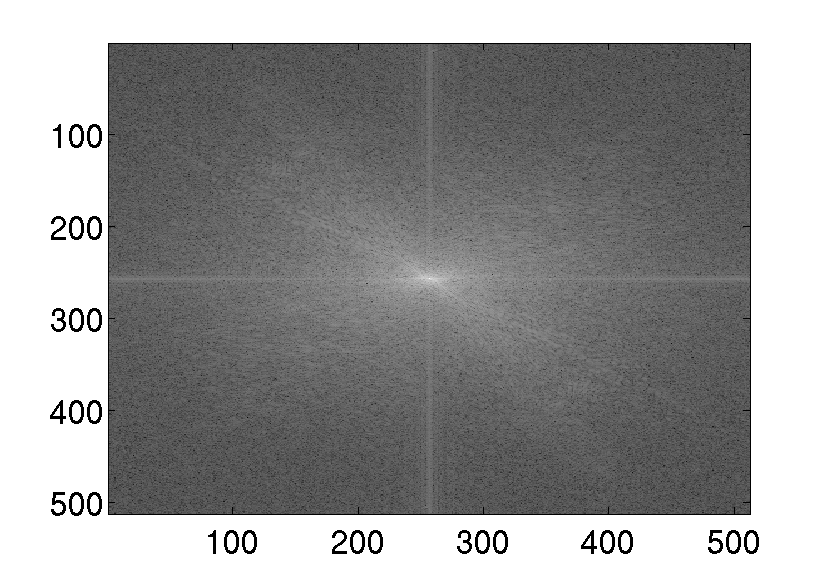}
	 \includegraphics[width=0.32\columnwidth]{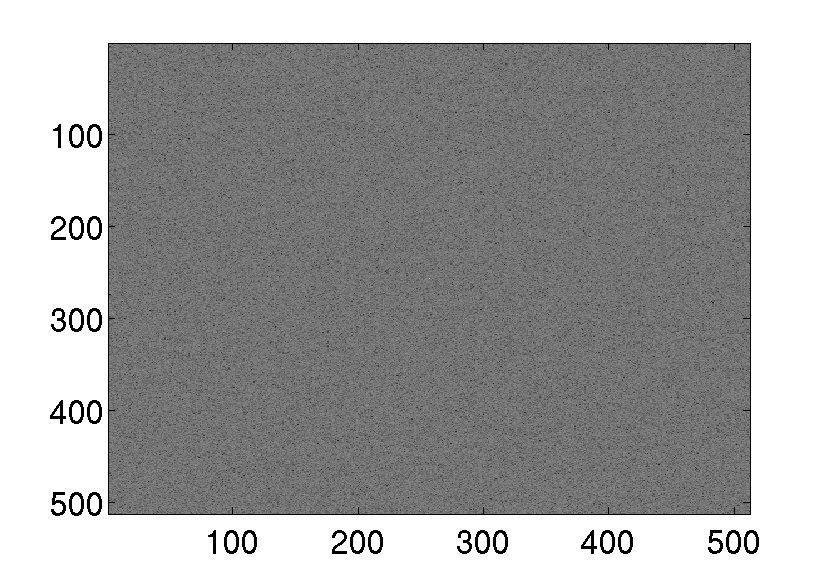} 
	 \includegraphics[width=0.32\columnwidth]{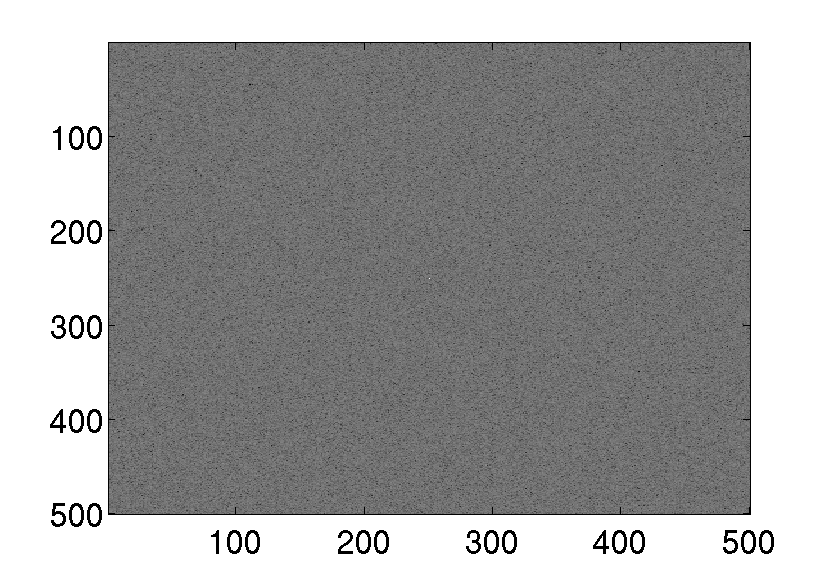}
        \\
         (g)\hspace{0.28\columnwidth}(h)\hspace{0.28\columnwidth}(i)
        \\
	\caption{Comparison of a plain image, cipher image and white noise. In the first row, the matrices analyzed are shown: (a) plain image, (b) cipher image and (c) white noise. 
	                In the second, the histogram analysis of the:  (d) plain image, (e) cipher image and (f) white noise. 
	                And in the last row, there is the Fourier power spectrum of: (g) plain image, (h) cipher image and (i) white noise.}
	\label{image}
\end{figure}

In the second experiment, the cipher was evaluated encrypting an image. Images are also appropriate to cryptoanalysis, as, they have strong global and local patterns and redundancy. 
The experiment is to compare three images: the original one, the cipher-image and white noise image. 
The frequency analysis has been performed and compared it to 2D Fourier power spectrum of the images (see Figure~\ref{image}). 
Notice that while the original image histogram has a lot of information concerning the image and both the cipher image and noise histograms are similar.
This fact shows how unintelligible the encrypted image is. 
Unlike the spectrum of the original image, the spectrum of the noise and the cipher-image does not present any  frequency information, which demonstrates that the information cannot be achieved in the cipher-image.         

\section{Fast and parallel version}
\label{fast}

Performance is the most important obstacle that prevents practical use of chaos based cryptography algorithms. 
In fact, they need float point and numerous calculations, which make the chaos cryptography programs much slower than the traditional ones.
To make the proposed algorithm feasible, we have improved its performance using parallel computation, which is in current use nowadays via the increase of multiple cores in the computer processor or with the use of graphical cards with multiple GPUs (Graphical Processing Unit).

The nature of our algorithm allows an intuitive parallelization. Consider a message of size $n$ and a machine with $p$ processors. 
The parallel algorithm splits the message into $p$ parts and executes each part into a different processor. 
For each process a particular initial condition is selected.

\begin{figure}[!htb]		
	\centering
	\includegraphics[width=0.9\columnwidth]{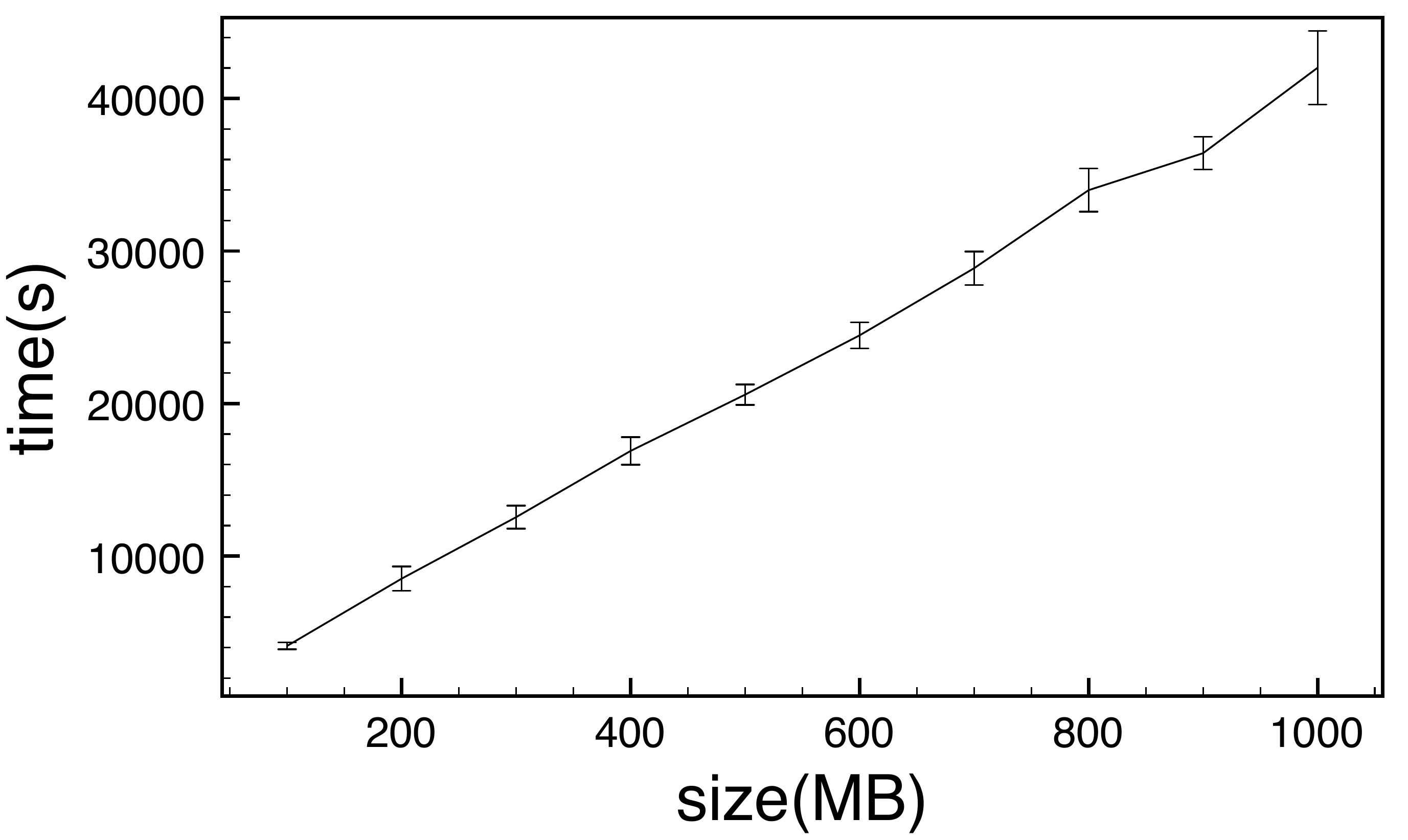} \\
	 \includegraphics[width=0.85\columnwidth]{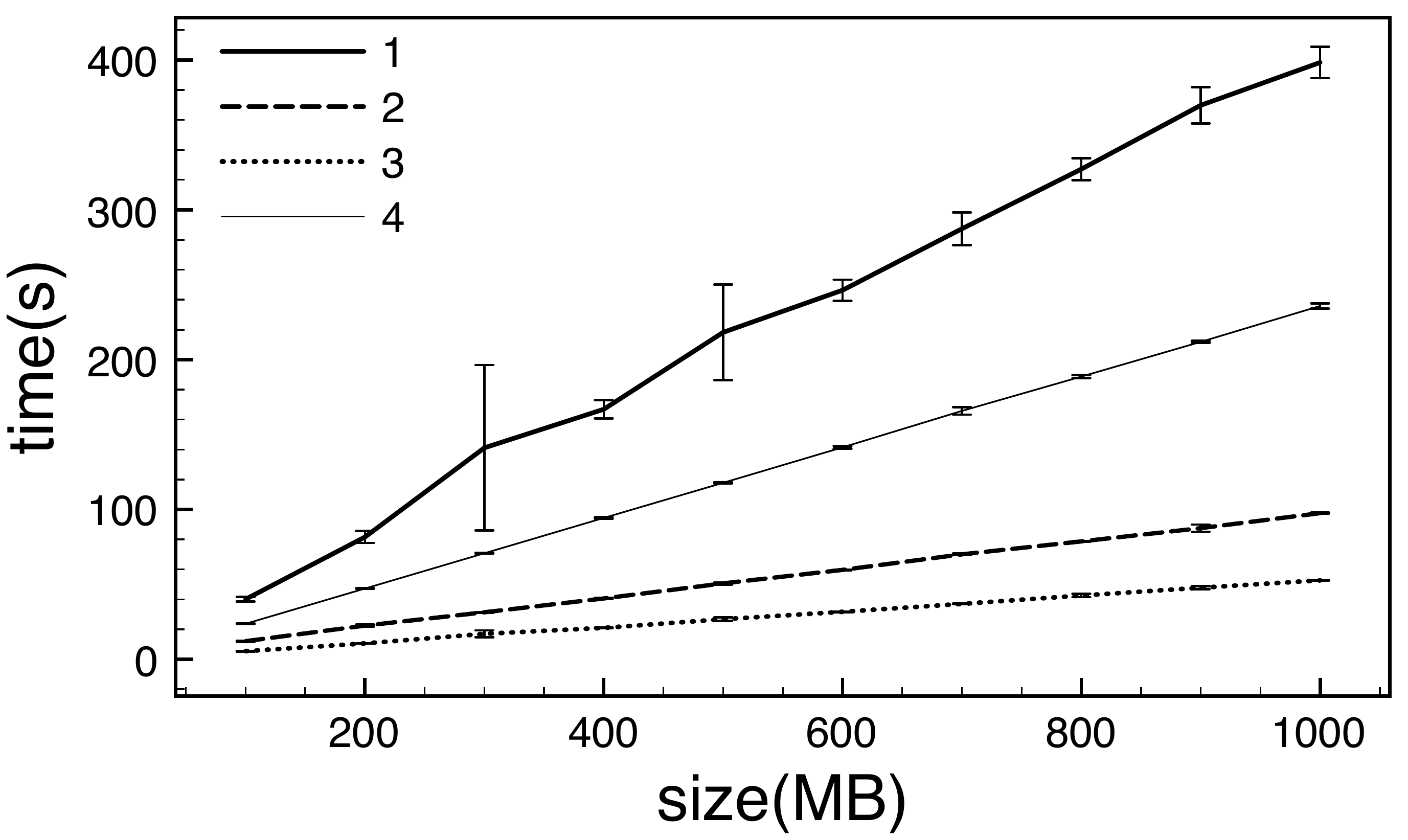} 
	\caption{Performance graphs. 
	(a) Sequential performance of the proposed algorithm size in MBytes versus time in seconds.
	(b) Comparison of the performance of the proposed algorithm in the GPU version and the AES cryptography method . 
	      Line 1 represents the performance of the proposed algorithm in a Nvidia geforce 8800 GT. 
	      Line 2 represents of the performance of the AES algorithm in a Intel Pentium 4 3.4 Ghz.
	      Line 3 represents of the performance of the AES algorithm in a Intel QuadCore Q8200 and line 4 represents the performance of the proposed algorithm in a Nvidia GTX 285.}
	\label{performance_graphs}
\end{figure}

The computer performance of the cryptography algorithm is an important item to make it feasible for real applications. 
The algorithm has been implemented in Java and also in C (fast version). 
The choice of Java was made due to its portability and capability to run on multiple operational systems and computer hardware (also small devices such as mobile phones). 
Two versions of the algorithm have been implemented: the full proposed method as described in Section~\ref{algorithm} and its parallel version described above.  
In the parallel version, the chaotic operation mode has been adapted to make the algorithm easy to be parallelized. 
This version is weaker than the original version, but it is faster and also highly parallelized. 
There is also a performance comparison of this implementation with the popular AES cryptography method~\cite{aes} shown in Figure~\ref{performance_graphs}. 

Figure~\ref{performance_graphs}a shows the plot of the time consumption versus the size of the file encrypted for the sequential strong version. 
The machine used was a Pentium 4 - 3.40 GHz, with 3 GB of memory and which runs Gentoo Linux distribution and a Intel QuadCore Q8200 with 4 GB of memory and which runs Debian Linux 5.0 distribution. 
The algorithm has an average of rate of approximately 1.5 MB/min. 
Note that the program was implemented in Java and can achieve a performance of 3 or 4 times faster if it is built using the C language. 
Although, this performance is very slow, it can be considered usable for high security purpose tasks. 

To improve the performance of the algorithm, a simpler (but still strong), fast and parallel version has been developed (Subsection~\ref{fast}). 
The fast parallel version was implemented in C and, to explore mass computer parallelism, the GPU (Graphics Processing Unit) architecture was adopted (8800 GT graphical card) and the CUDA (www.nvidia.com), a GPU programming tool for C, was used. 
Figure~\ref{performance_graphs}b presents the performance of the program compared with the AES. Both programs run on the same computer (described previously). 
The performance of the chaos based algorithm, which is approximately (approximate 2.5 MB/s in a Nvidia Geforce 8800GT and approximate 4.3 MB/s in a Nvidia GTX 285) is close to the AES (approximate 8.3 MB/s in a Intel Pentium 4 3.4 Ghz and approximate 18 MB/s in a Intel QuadCore Q8200). 
The great performance of the fast and parallel implementation allows immediate use of the chaos cryptography algorithm in real life applications, which has an advantage of being more secure then the traditional cryptography approach.
 
\section{Conclusion}
\label{conclusion}
In this paper, a new cryptograph algorithm based on the Lorenz's attractor and a novel method to carry out the operation mode with chaos (Chaotic Operation Mode) were proposed. 
The proposed algorithm in combination with the Chaotic operation mode achieves a strong cipher. 
The novel method has been implemented in Java (sequential strong version) and also in CUDA (parallel fast version) and experiments present the performance of cryptography and as well as the time of the algorithm has been presented and discussed. 
The performance of the method and the comparison with the AES algorithms demonstrate that the method is suitable and prepared for real life applications. 
The analysis of the algorithm and of the results, we notice that the algorithm is strong and perhaps very difficult to break into. 
However, it is very difficult to determine in fact, how hard t is to break into a cipher. 
Besides the set of experiments conducted and presented, a web version of the algorithm has been implemented. 
We would like to invite crypto-analysts to try to break into the proposed cipher, and help us to determine the reliability of the method and of course, to help develop of new methods. 
Cryptology is a dynamic science which is forever changing. 
The cryptography page can be found at http://www.mandelbrot.ifsc.usp.br/crypto\_lorenz, and the reader can find an interface for encode and decode messages and files.   

\section{Acknowledgements}
A. G. M. acknowledges support from CNPq (I. C. Grant).
A. S. M. acknowledges the support of CNPq (303990/2007-4, 476862/2007-8). 
O. M. B. would like to thank CNPq (306628/2007-4). 

%\bibliographystyle{apsrev}
%\bibliography{crypto}

\begin{thebibliography}{10}
\expandafter\ifx\csname url\endcsname\relax
  \def\url#1{\texttt{#1}}\fi
\expandafter\ifx\csname urlprefix\endcsname\relax\def\urlprefix{URL }\fi

\bibitem{serpent}
E.~Biham, R.~Anderson, L.~Knudsen, 
Serpent: A new block cipher proposal, in: 
In Fast Software Encryption '98, Springer-Verlag, 1998, pp. 222--238.

\bibitem{towfish}
B.~Schneier, J.~Kelsey, D.~Whiting, D.~Wagner, C.~Hall, N.~Ferguson, Twofish: A
  128-bit block cipher, in: in First Advanced Encryption Standard (AES)
  Conference, 1998.

\bibitem{idea}
X.~Lai, J.~L. Massey, A proposal for a new block encryption standard,
  Springer-Verlag, 1991, pp. 389--404.

\bibitem{aes}
J.~Daemen, V.~Rijmen, The Design of Rijndael: AES - The Advanced Encryption
  Standard, Springer-Verlag, 2002.

\bibitem{logaritimo}
M.~Diffie, W.~Hellman, 
New directions in cryptography, 
{\it Information Theory, IEEE Transactions} {\bf 22},  644 (1976).

\bibitem{RSA}
L.~A. R.~L.~Rivest, A.~Shamir, 
A method for obtaining digital signatures and public-key cryptosystems, 
{\it Commun. ACM} {\bf 21},  120 (1978).

\bibitem{curvas}
N.~Koblitz, 
Elliptic curve cryptosystems, 
{\it Mathematics of Computation} {\bf 48},  203 (1987).

\bibitem{crip_linear}
M.~Matsui, A.~Yamagishi, 
A new method for known plaintext attack of feal cipher, 
in: EUROCRYPT, 1992, pp. 81--91.

\bibitem{crip_diferenca}
E.~Biham, A.~Shamir, 
Differential cryptanalysis of the full 16-round des, 
in: CRYPTO '92: Proceedings of the 12th Annual International Cryptology
  Conference on Advances in Cryptology, Springer-Verlag, London, UK, 1993, pp.
  487--496.

\bibitem{criptoanalise1}
J.~P. Nicolas~Courtois, 
Cryptanalysis of block ciphers with overdefined systems of equations, 
{\it LNCS} {\bf 2501},  267 (2002).

\bibitem{wang_2008}
X.-Y. Wang and X.-J. Wang, 
Design of chaotic pseudo-random bit generator and its applications in stream-cipher cryptography, 
{\it Int. J. Mod. Phys. C} {\bf 19},  813 (2008).

\bibitem{Pecora:1990}
L.~Pecora, T.~Carroll, 
Synchronization in chaotic systems, 
{\it Phys. Rev. Lett.} {\bf 64},  821 (1990).

\bibitem{Chua}
L.~Kocarev, K.~S. Halle, K.~Eckert, L.~O. Chua, U.~Parlitz, 
Experimental demonstration of secure communications via chaotic synchronization,
{\it Int. J. Bifurcation and Chaos} {\bf 2}, 709 (1992).

\bibitem{Parlitz:1996p969}
U.~Parlitz, L.~Kocarev, T.~Stojanovski, H.~Preckel, 
Encoding messages using chaotic synchronization, 
{\it Phys. Rev. E} {\bf 53},  4351 (1996).

\bibitem{Bu:2004p1023}
S.~Bu, B.~Wang, 
Improving the security of chaotic encryption by using a simple modulating method, 
{\it Chaos Soliton and Fractals} {\bf 19},  919 (2004).

\bibitem{Murali}
K.~Murali, 
Heterogeneous chaotic systems based cryptography, 
{\it Phys. Lett. A} {\bf 272},  184 (2000).

\bibitem{Baptista}
M.~Baptista, 
Cryptography with chaos, 
{\it Phys. Lett. A} {\bf 240},  50 (1998).

\bibitem{Wong:2001}
W.-k. Wong, L.-p. Lee and K.-w. Wong,
A modified chaotic cryptographic method,
{\it Computer Physics Communications} {\bf 138},  234 (2001).

\bibitem{blocos}
T.~Xiang, X.~Liao, G.~Tang, Y.~Chen, K.~wo~Wong, 
A novel block cryptosystem based on iterating a chaotic map, 
{\it Phys. Lett. A} {\bf 349}, 109 (2005).

\bibitem{PRE}
R.~He, P.~G. Vaidya, 
Implementation of chaotic cryptography with chaotic synchronization, 
{\it Phys. Rev. E} {\bf 57},  1532 (1998).

\bibitem{AliPacha:2007}
A.~Ali-Pacha, N.~Hadj-Said, A.~M'Hamed, A.~Belgoraf, 
Lorenz's attractor applied to the stream cipher (ali-pacha generator), 
{\bf Chaos Soliton Fractals} {\bf 33},  1762 (2007).

\bibitem{Bell_Labs}
C.~Shannon, 
Communication theory of secrecy systems, 
{\it Bell System Technical  Journal}, {\bf 28}  656 (1949).

\bibitem{modo_operecao}
M.~Dworkin, 
Recommendation for block cipher modes of operation methods and techniques, 
NIST Special Publication 800-38A.

\bibitem{oliveira_2004}
G. M. B. Oliveira, A. R. Coelho and L. H. A. Monteiro, 
Cellular automata cryptographic model based on bi-directional toggle rules, 
{\it Int. J. Mod. Phys. C} {\bf 15},  1061 (2004). 

\bibitem{mora_2002}
J. C. S. T. Mora,  
Transitive behavior in reversible one-dimensional cellular automata with a Welch index 1, 
{\it Int. J. Mod. Phys. C} {\bf 13},  837 (2002). 

\bibitem{mora_2003}
J. C. S. T. Mora, S. V. C. Vergara, G. J. Mart\'{\i}nez and H. V. McIntosh, 
Spectral properties of reversible one-dimensional cellular automata, 
{\it Int. J. Mod. Phys. C} {\bf 14},   379 (2003). 

\bibitem{hernandez_2003}
J. C. S. T. Mora, M. G. Hern\'andez,  G. J. Mart\'{\i}nez and S. V. C. Vergara, 
Extensions in reversible one-dimensional cellular automata are equivalent with the full shift, 
{\it Int. J. Mod. Phys. C} {\bf 14},  1143 (2003). 


\bibitem{delrey_2006} 
A. M. del Rey and G. R. S\'anchez, 
On the reversibility of 150 Wolfram cellular automata, 
{\it Int. J. Mod. Phys. C} {\bf 17},  975 (2006). 

\bibitem{delrey_2009} 
A. M. del Rey and G. R. S\'anchez, 
Reversibility of a symmetric linear cellular automata, 
{\it Int. J. Mod. Phys. C}  {\bf 20},   1081-1086 (2009). 

\bibitem{Arroyo}
X.-Y. Wang, C.-F. Duan and N. Gu, 
A new chaotic cryptography based on ergodicity, 
{\it Int. J. Mod. Phys. B} {\bf 22},  901 (2008). 

\bibitem{martinez_2009}
A. S. Martinez, R. S. Gonz\'alez and A. L. Espindola, 
Generalized exponential function and discrete growth models
{Physica A} {\bf 388}, 2922–2930 (2009). 
\end{thebibliography}

\end{document}